\documentclass[aps,floats]{revtex4}
\usepackage{amsmath,amssymb}
\usepackage{graphicx,epsfig}

\begin{document}
\bibliographystyle {plain}

\def\oppropto{\mathop{\propto}} 
\def\opmin{\mathop{\min}} 
\def\opmax{\mathop{\max}} 
\def\opsimeq{\mathop{\simeq}}
\def\opoverderline{\mathop{\overline}}
\def\operarrow{\mathop{\longrightarrow}}
\def\opsim{\mathop{\sim}}

\def\fig#1#2{\includegraphics[height=#1]{#2}}
\def\figx#1#2{\includegraphics[width=#1]{#2}}


\title{ Statistical properties of two-particle transmission at Anderson transition } 


 \author{ C\'ecile Monthus and Thomas Garel }
\affiliation{Institut de Physique Th\'{e}orique, CNRS and CEA Saclay
 91191 Gif-sur-Yvette cedex, France}

\begin{abstract}
The ensemble of $L \times L$ power-law random banded matrices, where the random hopping $H_{i,j}$ decays as a power-law $(b/\vert i-j \vert)^a$, is known to present an Anderson localization transition at $a=1$, where one-particle eigenfunctions are multifractal. Here we study numerically, at this critical point, the statistical properties of the transmission $T_2$ for two distinguishable particles, two bosons or two fermions. We find that the statistics of $T_2$ is multifractal, i.e. the probability to have $T_2(L) \sim 1/L^{\kappa}$ behaves as $L^{\Phi_2(\kappa)}$, where the multifractal spectrum $\Phi_2(\kappa)$ for fermions is different from the common multifractal spectrum concerning distinguishable particles and bosons. However in the three cases, the typical transmission $T_2^{typ}(L)$ is governed by the same exponent $\kappa_2^{typ}$, which is much smaller than the naive expectation $2\kappa_1^{typ}$, where $\kappa_1^{typ}$ is the typical exponent of the one-particle transmission $T_1(L)$.

\end{abstract}

\maketitle

\section{ Introduction }

Whereas Anderson localization phenomena \cite{anderson} are rather well
understood for a single particle
 (see the reviews \cite{janssenrevue,markos,mirlinrevue}),
the case of interacting particles in a random potential
 has remained much more challenging (see the review \cite{belitz}
and more recent works \cite{manybodyloc,
levitov,silvestrov,gornyi,huse,fleury}).
Since the case of a finite density of particles can be studied
numerically only for small system sizes,
it is natural to consider first the simpler case of only
two interacting particles (T.I.P.) in a random potential.
In dimension $d=1$, where the one-particle model is always in the localized
phase with some localization length $\lambda_1$, it has been found 
that the T.I.P. is also always localized, but with a localization
length $\lambda_2$ that may become much larger than $\lambda_1$
\cite{dorokhov,shepel,pichard,felix,ponomarev,
song,waintal,vonOppen,leadbeater,schreiber}.
In dimension $d=2$, where the one-particle model is again always in the localized
phase, the possibility of a delocalization transition has been studied
for short-range interaction \cite{ortuno}
and for Coulomb interaction \cite{shepel2D,talamentes}.

In the present paper, we are interested in the two-particle 
transport properties at  
the Anderson localization transition of the one-particle problem,
where the one-particle eigenstates are multifractal 
\cite{janssenrevue,mirlinrevue}. We are not aware of previous studies
on this question (see however \cite{quasiperiodic} concerning
the quasi-periodic Aubry-Andr\'e transition).
Since for the tight-binding model in dimension $d=3$ where there
exists an Anderson transition, the two-particle model cannot
 be studied numerically for large enough system sizes and large enough
statistics on the disordered samples to obtain accurate results,
we have chosen to focus here on
the Power-law random banded matrices (PRBM) model,
and to study numerically the statistical properties 
of the two-particle transmission $T_2$.
 
The paper is organized as follows.
In section \ref{model}, we recall the
 Power-law random banded matrices (PRBM) model 
and introduce the observables that characterize transport properties
for the two-particle model.
In section \ref{numet2}, we describe our numerical results concerning 
the statistical properties of the  
 transmission $T_2$ for two distinguishable particles, 
two bosons and two fermions.
Our conclusions are summarized in section \ref{conclusion}.
In Appendix \ref{multif1pener}, we describe our numerical results concerning the multifractal properties of the one-particle model as a function of the energy $E$, which turn out to be useful to understand the statistics of $T_2$
 discussed in the text.

\section{ Model and observables }

\label{model}

\subsection{ Reminder on the Power-law random banded matrices (PRBM) model  }

Beside the usual short-range Anderson tight-binding model
in finite dimension $d$, other models displaying Anderson localization
have been studied,
in particular the Power-law Random Banded Matrix (PRBM) model,
which can be viewed as a one-dimensional model with long-ranged
random hopping decaying as a power-law $(b/r)^a$ of the distance $r$
with exponent $a$ and parameter $b$.
The Anderson transition at $a=1$ between localized ($a>1$)
 and extended ($a<1$) states
has been characterized in \cite{mirlin96} via a mapping onto a non-linear
sigma-model. The properties of the critical points at $a=1$ 
have been then much studied, in particular
the statistics of eigenvalues \cite{varga00,kra06,garcia06}, 
and the multifractality of eigenfunctions 
\cite{mirlin_evers,cuevas01,cuevas01bis,varga02,cuevas03,mirlin06},
including boundary multifractality \cite{mildenberger}.

\begin{figure}[htbp]
 \includegraphics[height=6cm]{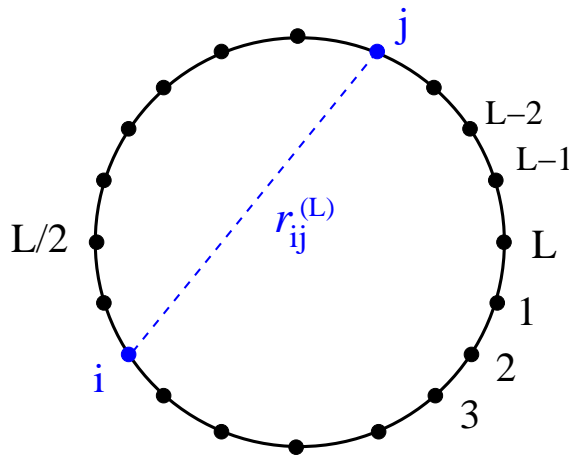}
\vspace{1cm}
\caption{ (Color on line) 
The ensemble of power-law random banded matrices of size 
$L \times L$ can be represented as a ring of $L$ sites,
where the matrix element $H_{i,j}$ between the sites $i$ and $j$
is a Gaussian variable of zero-mean $\overline{H_{i,j}}=0$
 and of variance given by Eq. \ref{defab} in terms of the distance 
 $r_{i,j}$ of Eq. \ref{rijcyclic}. 
  }
\label{ring}
\end{figure}

More precisely, we consider here the model shown on Fig \ref{ring},
with $L$ sites $i=1,2,..L$ in a ring geometry
with periodic boundary conditions.
The appropriate distance $r_{i,j}$
between the sites $i$ and $j$ is defined as \cite{mirlin_evers}
\begin{eqnarray}
r_{i,j}^{(L)} = \frac{L}{\pi} \sin \left( \frac{ \pi (i-j) }{L} \right)
\label{rijcyclic}
\end{eqnarray}
The ensemble of power-law random banded matrices of size 
$L \times L$ is then defined as follows : 
  the matrix elements $H^{(1)}_{i,j}$ are independent Gaussian
variables of zero-mean $\overline{H^{(1)}_{i,j}}=0$ and of variance
\begin{eqnarray}
\overline{ (H^{(1)}_{i,j})^2 } = \frac{1}{1+ \left( \frac{r_{i,j}}{b}\right)^{2a}}
\label{defab}
\end{eqnarray}
The most important properties of this model are the following.
The value of the exponent $a$ determines the localization properties
\cite{mirlin96} : 
 for $a>1$ states are localized with integrable power-law tails,
whereas for $a<1$ states are delocalized.
At criticality $a=1$, states become multifractal 
\cite{mirlin_evers,cuevas01,cuevas01bis,varga02} and exponents
depend continuously of the parameter $b$, which plays a role analog
to the dimension $d$ in short-range Anderson transitions 
\cite{mirlin_evers} : the limit $b \gg 1$ corresponds to
weak multifractality ( analogous to the case $d=2+\epsilon$)
and can be studied via the mapping onto a non-linear sigma-model
 \cite{mirlin96},
whereas the case $b \ll 1$ corresponds to strong multifractality
( analogous to the case of high dimension $d$)
and can be studied via Levitov renormalization \cite{levitovRG,mirlin_evers}.
Other values of $b$ have been studied numerically  
\cite{mirlin_evers,cuevas01,cuevas01bis,varga02}.
The statistical properties of the Landauer transmission
for a single particle between the opposite points $L/2$ and $L$ 
has been studied in detail in our previous work 
\cite{us_twopoints}
(results concerning other scattering geometries can be found in 
\cite{us_many}).

\subsection{ Transmission of two distinguishable
particles, two bosons, or two fermions }

In this paper, we consider the two-particle model defined by the Hamiltonian
\begin{eqnarray}
H^{(2)} = H^{(1)} \otimes 1 + 1  \otimes H^{(1)} 
\label{H2}
\end{eqnarray}
As stressed in \cite{felix,song,vonOppen,leadbeater,schreiber}
for the one-dimensional T.I.P. model, the important observable
to characterize the two-particle transport properties is
the Green function 
\begin{eqnarray}
G_{E_2} \equiv \frac{1}{H^{(2)}-E_2} 
\end{eqnarray}
between doubly occupied sites along the diagonal $r_1=r_2$.
In our present notations concerning the P.R.B.M. model (see Fig. \ref{ring}),
we will thus focus on the transmission
\begin{eqnarray}
T_2 \equiv \vert < \frac{L}{2}, \frac{L}{2} \vert G_{E_2=0} \vert L,L> \vert^2
\label{t2}
\end{eqnarray}
at zero energy $E_2=0$ (center of the band).
It is important to stress that even if there
 is no explicit interaction in the Hamiltonian
of Eq. \ref{H2}, the two-particle Green function cannot be factorized
into one-particle properties \cite{song,leadbeater}.
We will indeed find below non-trivial properties for $T_2$.
As a comparison, one may also consider 
the transmission of one of the two particles with the other held fixed
 (see Eq. 7 of \cite{vonOppen})
\begin{eqnarray}
T_{2,(1f)}  \equiv \vert < \frac{L}{2},L \vert G \vert L,L> \vert^2 
\label{t2(1f)}
\end{eqnarray}

\section{ Numerical results on the statistical properties of $T_2$ }

\label{numet2}

\subsection{ Numerical procedure }

We have used 
an exact diagonalization method the one-particle P.R.B.M. model
 $H^{(1)}$ 
for the localization transition critical value $a=1$ and for the
parameter $b=0.1$ (Eq. \ref{defab}).
For each disordered sample, we note $e_n$
the $L$ eigenenergies ($n=1,2,..L$)  and $\phi_n(i)$  
the corresponding normalized eigenstates ($i=1,2,..L$) 
\begin{eqnarray}
H^{(1)}  \phi_n(i) = e_n \phi_n(i)
\label{diagoH1}
\end{eqnarray}

To compute the two-particle Green function, 
we now have to know the symmetry properties
with respect to the exchange of the two particles.

\subsubsection{Two distinguishable particles (No symmetry conditions)}

For two distinguishable particles, an orthonormal basis of eigenstates 
of $H^{(2)}$ is given
by the following $L^2$ states labelled by two integers 
 $1 \leq n \leq L$ and $1 \leq m \leq L$ 
\begin{eqnarray}
\psi_{n,m} (i,j) = \phi_n(i) \phi_m(j)
\label{diago2Partpsi}
\end{eqnarray}
of energy
\begin{eqnarray}
E_{n,m}  = e_n+e_m
\label{diago2Parte}
\end{eqnarray}

The two-particle Green function at zero energy $E_2=0$ then reads
\begin{eqnarray}
G_{E_2=0} (i,j ; i',j')
&&  = - \sum_{n=1}^L \sum_{m=1}^L
\frac{\psi_{n,m}^* (i,j) \psi_{n,m} (i',j')}{E_{n,m}} \\
&& = - \sum_{n=1}^L \sum_{m=1}^L
\frac{\phi_n^*(i) \phi_m^*(j) \phi_n(i') \phi_m(j')}{e_n+e_m}
\label{green2Part}
\end{eqnarray}

\subsubsection{Two  Bosons (Symmetry condition) }

An orthonormal basis of eigenstates is given
by the following $L(L+1)/2$ symmetric states labelled by two integers 
  $1 \leq n \leq m \leq L$ 
\begin{eqnarray}
\psi^B_{n,n} (i,j) && = \phi_n(i) \phi_n(j) \\
\psi^B_{n,m} (i,j) && =
 \frac{\phi_n(i) \phi_m(j)+\phi_m(i) \phi_n(j) }{\sqrt 2}
\label{diago2bosonspsi}
\end{eqnarray}
of energy given by Eq. \ref{diago2Parte}.

The two-boson Green function at zero energy $E_2=0$ then reads
\begin{eqnarray}
G_{E_2=0}^B(i,j ; i'j')
  = - \sum_{n=1}^{L} \sum_{m=n}^L
\frac{\psi_{n,m}^{B*} (i,j) \psi_{n,m}^B (i',j')}{e_n+e_m} 
\label{green2bosons}
\end{eqnarray}

\subsubsection{Two  Fermions (Antisymmetry condition)}

An orthonormal basis of eigenstates is given
by the following $L(L-1)/2$ antisymmetric states labelled by two integers 
  $1 \leq n < m \leq L$ 
\begin{eqnarray}
\psi^F_{n,m} (i,j) = \frac{\phi_n(i) \phi_m(j)-\phi_m(i) \phi_n(j) }{\sqrt 2}
\label{diago2fermionspsi}
\end{eqnarray}
of energy given by Eq. \ref{diago2Parte}

The two-fermion Green function at zero energy $E_2=0$ then reads
\begin{eqnarray}
G_{E_2=0}^F(i,j ; i'j')
  = - \sum_{n=1}^{L-1} \sum_{m=n+1}^L
\frac{\psi_{n,m}^{F*} (i,j) \psi_{n,m}^F (i',j')}{e_n+e_m} 
\label{green2fermions}
\end{eqnarray}

For fermions where double occupancy is forbidden, 
we have modified the definitions of Eqs \ref{t2} and \ref{t2(1f)}
for the transmissions into 
\begin{eqnarray}
T_2^F \equiv \vert < L/2,L/2-1 \vert G \vert L,L-1> \vert^2
\label{t2F}
\end{eqnarray}
and
\begin{eqnarray}
T_{2,(1f)}^F \equiv \vert < L/2-1,L \vert G \vert L-1,L> \vert^2 
\label{t2(1f)F}
\end{eqnarray}

The results given below correspond to sizes
$50 \leq L \leq 2000$,
with corresponding statistics of $5.10^7 \geq n_s(L) \geq 1150$
 independent samples. To improve the statistics, we have considered,
for each disordered sample, the transmission $T_2$ between the $L/2$ pairs
of opposite points.
All results concern the zero-energy ($E_2=0$) transmission $T_2$
at the critical point $a=1$ and the value $b=0.1$ (see
Eq. \ref{defab}).
 We first focus
on the scaling of the typical transmission before
we turn to the multifractal spectrum.

\subsection{ Typical transmission $T_2^{typ}(L)$ as a function of $L$ }

\begin{figure}[htbp]
 \includegraphics[height=8cm]{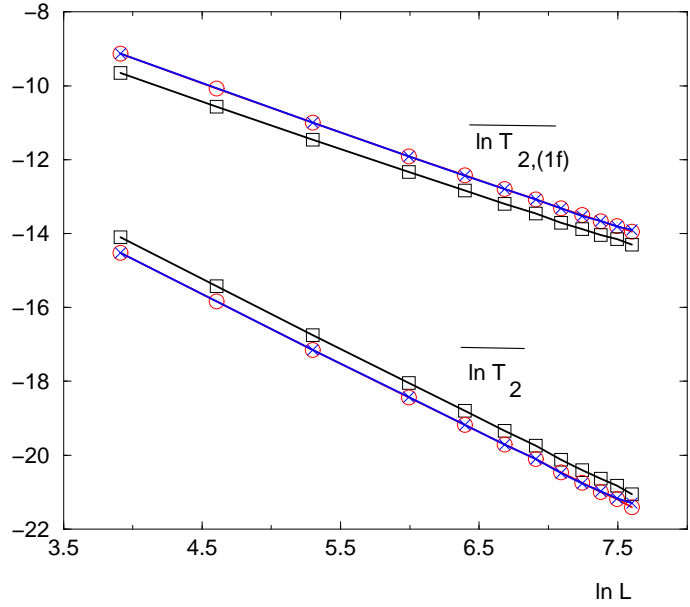}
\vspace{1cm}
\caption{ (Color on line) 
Scaling of the typical two-particle transmission 
$T_2^{typ}$ at criticality $a=1$ for $b=0.1$ and at zero-energy $E=0$ :
 $ \ln T_2^{typ}(L) \equiv  \overline{ \ln T_2(L) } $
as a function of $ \ln L$ for 
fermions ($\square$), bosons ($\bigcirc$) and distinguishable
 particles ($\times$)
yield the same exponent $\kappa_2^{typ} \simeq 1.86$ (Eq. \ref{kappa2typ}).
As a comparison, the scaling of the typical transmission $T_{2,(1f)}^{typ}$ 
of Eq. \ref{t2(1f)} (representing the transmission
of one of the two particles with the other held fixed) 
corresponds to the exponent $\kappa_{2,(1f)}^{typ} \simeq 1.3$
 ( Eq. \ref{kappa1typ}). }
\label{figt2typ}
\end{figure}

We find that the typical two-particle transmission
\begin{eqnarray}
   T^{typ}_2(L) \equiv e^{ \overline{ \ln T_2(L) } } 
\label{defTtyp}
\end{eqnarray}
decays as the power-law
\begin{eqnarray}
   T^{typ}_2(L) 
\oppropto_{L \to \infty} \frac{1}{L^{\kappa_2^{typ}}}
\label{defTtypcriti}
\end{eqnarray}
where
\begin{eqnarray}
 \kappa_2^{typ} \simeq 1.86
\label{kappa2typ}
\end{eqnarray}
is the same for two distinguishable particles, two bosons or two fermions
as shown on Fig. \ref{figt2typ}.

As a comparison, we also show on Fig. \ref {figt2typ}
the typical transmission $T_{2,(1f)}^{typ}$ of Eq. \ref{t2(1f)}
representing the transmission
of one of the two particles with the other held fixed :
  for distinguishable particles, bosons or fermions, it 
is governed by the same exponent
\begin{eqnarray}
 \kappa_{2,(1f)}^{typ} \simeq 1.3
\label{kappa1typ}
\end{eqnarray}
that coincides, within our error bars, with the exponent $\kappa_1^{typ}$
 measured in \cite{us_twopoints} for the one-particle model.

\subsection{ Multifractal statistics of $T_2$  }

\begin{figure}[htbp]
 \includegraphics[height=8cm]{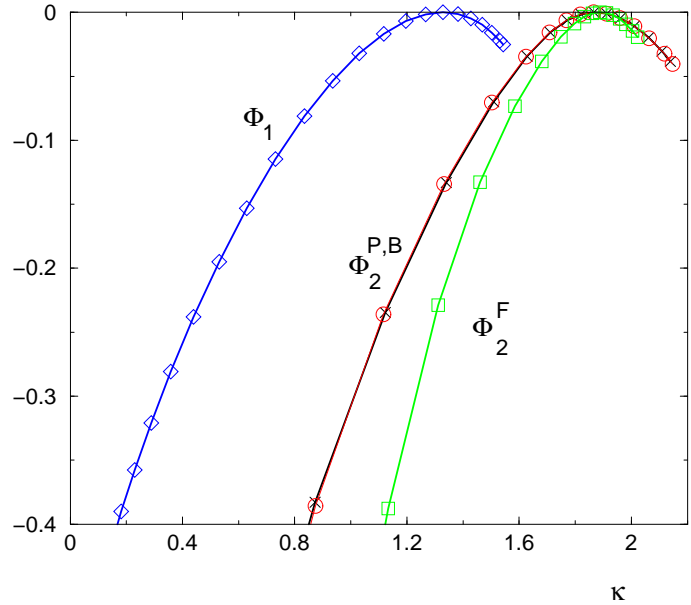}
\vspace{1cm}
\caption{ (Color on line) 
Multifractal spectra $\Phi_2(\kappa)$ describing the statistics of
the transmission $T_2$ at criticality $a=1$ for $b=0.1$ and at zero-energy $E=0$ :  for two fermions ($\square$), two bosons ($\bigcirc$) and two distinguishable
 particles ($\times$).
As a comparison, the multifractal spectrum $\Phi_1(\kappa)$
describing the statistics of the corresponding one-particle transmission 
is shown ($\diamond$). }
\label{figphi2kappa}
\end{figure}

We find that the statistics of $T_2$ is multifractal,
 i.e. that the probability to have $T_2(L) \sim 1/L^{\kappa}$
 behaves as
\begin{equation}
{\rm Prob}\left( T_2(L) \sim L^{-\kappa}  \right) dT
\oppropto_{L \to \infty} L^{\Phi_2(\kappa) } d\kappa
\label{phi2kappa}
\end{equation} 
We show on Fig. \ref{figphi2kappa}
 the multifractal spectra $\Phi_2(\kappa)$
corresponding to two distinguishable particles, two bosons
and two fermions.
We find that the spectra for distinguishable particles and bosons
coincide, whereas the spectrum for fermions is clearly distinct,
except around the maximum $\Phi_2(\kappa_2^{typ})=0$ 
associated to the same typical value $\kappa_2^{typ}$ of Eq. \ref{kappa2typ}.
This can be explained as follows :
the transmission $T_2$ for distinguishable particles and bosons both involve 
coinciding points (Eq. \ref{t2}), whereas the transmission $T_2$ 
for fermions involves neighboring points (Eq. \ref{t2F}).
Besides their common typical scaling, one thus expects differences
in their statistics.

As a comparison, we also show on Fig. \ref{figphi2kappa}
the multifractal spectrum $\Phi_1(\kappa)$
describing the statistics of the corresponding one-particle transmission $T_1$.
A natural question is of course whether the multifractal spectrum  
$\Phi_2(\kappa)$ 
can be related to $\Phi_1(\kappa)$
or to the singularity spectrum of one-particle eigenfunctions.

\subsection{ Discussion : relation with the statistics of 
one-particle eigenfunctions }

We first recall the case of the one-particle transmission,
before we turn to the analysis of $T_2$

\subsubsection{Analysis of the one-particle transmission in terms of one-particle eigenfunctions }

In terms of the energies $e_n$ and eigenfunctions $\phi_n$ of the one-particle model (Eq. \ref{diagoH1}), the one-particle zero-energy Green function reads
\begin{eqnarray}
g_{E_1=0} (i ; i')
 = - \sum_{n=1}^L \frac{\phi_n^*(i) \phi_n(i') }{e_n}
\label{green1Part}
\end{eqnarray}
In the limit of large size $L$ where the levels become dense,
the zero-energy Green function of Eq. \ref{green2Part}
become
\begin{eqnarray}
g_{E_1=0} (i ; i')
 \sim - L^d \int de \rho(e) \frac{\phi_{e}^*(i) \phi_{e}(i') }{e}
\label{green1Partconti}
\end{eqnarray}
which is dominated by the neighborhood of $e=0$
\begin{eqnarray}
g_{E_1=0} (i ; i')
 \sim - L^d \phi_{e=0}^*(i) \phi_{e=0}(i') 
\label{green1Partapprox}
\end{eqnarray}
so that the one-particle transmission scales as
\begin{eqnarray}
T_1 (i,i') = \vert g_{E_1=0} (i ; i') \vert^2
 \sim L^{2d} \vert \phi_{e=0}(i)\vert^2
 \vert \phi_{e=0}(i')\vert^2 
\label{trans1pres}
\end{eqnarray}
When the distance $\vert i-i' \vert$ is of the order of the system size $L$,
the weights $\vert \phi_{e=0}(i)\vert^2 $ and $\vert \phi_{e=0}(i')\vert^2 $
can be considered as independent. Then the multifractal spectrum $\Phi_1(\kappa)$
describing the distribution of the one-point transmission
\begin{equation}
{\rm Prob}\left( T_1 \sim L^{-\kappa}  \right) dT
\oppropto_{L \to \infty} L^{\Phi_1(\kappa) } d\kappa
\label{phi1kappa}
\end{equation}
can be written as (here with $d=1$)
\begin{eqnarray}
\Phi_1(\kappa \geq 0) = 2 \left[ f( \alpha= d+ \frac{\kappa}{2}   ) -d \right]
\label{resphikappa1}
\end{eqnarray}
in terms of the singularity spectrum $f(\alpha)$ of zero-energy eigenfunctions
(see more details in \cite{janssen99,us_twopoints}).

\subsubsection{Analysis of the two-particle transmission 
in terms of one-particle eigenfunctions }

We now try to analyse the two-point transmission $T_2$
 for two distinguishable particles
along the same lines.
In the limit of large size $L$ where the levels become dense, 
the zero-energy Green function of Eq. \ref{green2Part}
become
\begin{eqnarray}
G_{E_2=0} (i,i ; i',i')
\sim - L^{2d} \int de \rho(e)  \int de' \rho(e') 
\frac{\phi_{e}^*(i) \phi_{e'}^*(i) \phi_{e}(i') \phi_{e'}(i')}{e+e'}
\label{green2Partconti}
\end{eqnarray}
which is dominated by the region $e+e'=0$
\begin{eqnarray}
G_{E_2=0} (i ; i')
\sim - L^{2d} \int de \rho^2(e)  
\phi_{e}^*(i) \phi_{-e}^*(i) \phi_{e}(i') \phi_{-e}(i')
\label{green2Partapprox}
\end{eqnarray}
(where we have used the symmetry $\rho(e)=\rho(-e)$ 
around the center of the band $e=0$ for the one-particle density
of states).
The two-particle transmission is then expected to scale as
\begin{eqnarray}
T_2 (i,i') = \vert G_{E_2=0} (i ; i') \vert^2
 \sim L^{4d} \int de \rho^2(e)  \int de' \rho^2(e') 
\left[ \phi_{e}^*(i) \phi_{-e}^*(i) \phi_{e'}^*(i) \phi_{-e'}^*(i) \right]
\times \left[ \phi_{e}(i') \phi_{-e}(i')  \phi_{e'}(i') \phi_{-e'}(i')  \right]
\label{trans2pcorre}
\end{eqnarray}
So we do not expect any simple expression
 for the multifractal spectrum $\Phi_2(\kappa)$ :
firstly, $T_2$ contains eigenfunctions of any energy $e$,
and the singularity spectrum $f(\alpha)$ of one-particle eigenfunctions
depends continuously on the energy $e$
 (see more details in Appendix \ref{multif1pener}); 
 secondly, $T_2$ involves complicated correlations of eigenfunctions
of various energies (studies of two-eigenfunctions correlations can
be found in \cite{correFM,correCK}).

It is however natural to consider the simplest approximation :
 if the integrals in Eq. \ref{trans2pcorre} were dominated by $e=0=e'$,
one would obtain a direct relation with the one-particle transmission
of Eq. \ref{trans1pres}
\begin{eqnarray}
T_2^{{\rm approx}(a)} (i,i') 
 \sim L^{4d}  \vert \phi_{e=0}(i)\vert^4 \vert \phi_{e=0}(i')\vert^4 
\sim ( T_1 (i,i'))^2
\label{trans2approxa}
\end{eqnarray}
In particular, the typical exponent $\kappa_2^{typ}$ would read
\begin{eqnarray}
  \kappa_2^{typ} = 2 \kappa_1^{typ}
\label{kappa2typapproxa}
\end{eqnarray}
Our numerical results described above (Eqs \ref{kappa2typ} and \ref{kappa1typ}) 
show that $\kappa_2^{typ}$ is in fact much smaller than $(2 \kappa_1^{typ})$.
Our conclusion is thus that this simple approximation 
is very bad, and that correlations between 
one-particle eigenfunctions at various energies play a major role in
the two-particle transmission $T_2$.

 \section{ Conclusion}

\label{conclusion}

In this paper, we have studied numerically the statistical properties
of the two-particle transmission $T_2(L)$ at the critical point
of the PRBM model where one-particle eigenfunctions are known to be
multifractal. Our conclusion is that $T_2(L)$ is multifractal
i.e. the probability to have $T_2(L) \sim 1/L^{\kappa}$
 behaves as $L^{\Phi_2(\kappa)}$, 
where the multifractal spectrum $\Phi_2(\kappa)$
for fermions is different from the common multifractal
spectrum concerning distinguishable particles and bosons,
because the double occupancy of a single site and
the occupancy of two neighboring sites have different statistics
at criticality. However in the three cases, the typical 
transmission $T_2^{typ}(L)$ is governed by the same exponent $\kappa_2^{typ}$,
which is much smaller than the naive expectation $2\kappa_1^{typ}$,
where $\kappa_1^{typ}$ is the typical exponent 
of the one-particle transmission $T_1(L)$.
This suggests that $T_2(L)$ probes non-trivial 
correlations of one-particle eigenfunctions of various energies.

 \appendix

 \section{ Multifractal statistics of the one-particle transmission 
 as a function of the energy}

\label{multif1pener}

\begin{figure}[htbp]
 \includegraphics[height=8cm]{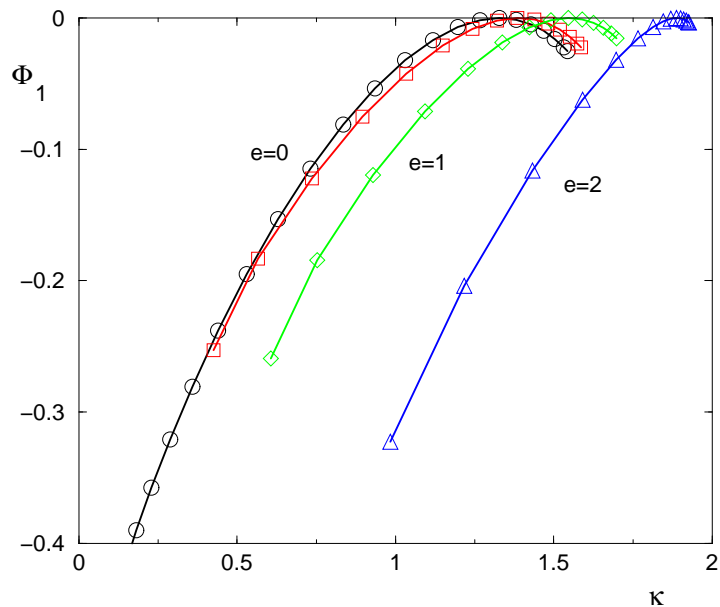}
\vspace{1cm}
\caption{ (Color on line) 
The multifractal spectrum $\Phi_1(\kappa)$ describing the statistics
of the one-particle transmission (Eq. \ref{phi1kappa}) at criticality
$a=1$ and $b=0.1$ for four values of the energy :
$e=0$ ($\bigcirc$), $e=0.5$ ($\square$), $e=1$ ($\diamond$) and $e=2$ ($\triangle$). 
  }
\label{figmultif1pener}
\end{figure}

As first discussed in \cite{janssen99}
for the special case of the two dimensional quantum Hall transition, the 
critical probability distribution of the one-particle transmission $T_1$ 
at an Anderson transition critical point 
 takes
 the form of Eq. \ref{phi1kappa} where
the multifractal spectrum $\Phi_1(\kappa)$ can be related to the 
singularity spectrum $f(\alpha)$ of critical eigenstates
 via Eq. \ref{resphikappa1}.

For the P.R.B.M. model, numerical results on $\Phi_1(\kappa)$
can be found in \cite{us_twopoints} at the critical point $a=1$
and at zero energy $e=0$
for various values of the parameter $b$ (Eq. \ref{defab}).

Here we show on Fig. \ref{figmultif1pener}
how the multifractal spectrum $\Phi_1(\kappa)$
at criticality $a=1$ for the value $b=0.1$ 
changes as a function of the energy $e$.
In particular, the corresponding typical values read
\begin{eqnarray}
 \kappa_1^{typ}(e=0) && \simeq 1.33 \nonumber \\
 \kappa_1^{typ}(e=0.5) && \simeq 1.38 \nonumber \\
 \kappa_1^{typ}(e=1) && \simeq 1.55 \nonumber \\
 \kappa_1^{typ}(e=2) && \simeq 1.89 
\label{kappa1typener}
\end{eqnarray}
Via Eq. \ref{resphikappa1}, 
this shows that the singularity spectrum $f(\alpha)$
of critical eigenfunctions changes with the energy $e$.
(The dependence on $e$ of $f(\alpha)$ has been studied in \cite{hallE}
for quantum Hall wavefunctions
as a function of the Landau level).

Since the zero-energy two-particle transmission $T_2$ of Eq. \ref{trans2pcorre}
contains one-particle eigenfunctions of various energies, 
that are characterized
by different multifractal singularity spectra, we do not expect any simple
expression for the multifractal spectrum $\Phi_2(\kappa)$ of $T_2$.

\end{document}